\documentclass[aps,prb,twocolumn,showpacs,amsmath,amssymb]{revtex4}
\usepackage{graphicx}
\usepackage{bm}
\def\l{\langle\!\langle}
\def\r{\rangle\!\rangle}
\def\bea{\begin{eqnarray}}
\def\eea{\end{eqnarray}}

\begin{document}

\title{Theory of quantum noise detectors based on resonant tunneling}
\author{Eugene V. Sukhorukov}
\email{Eugene.Sukhorukov@physics.unige.ch}
\author{Jonathan Edwards}
\affiliation{D\'epartement de Physique Th\'eorique, Universit\'e de Gen\`eve,
CH-1211 Gen\`eve 4, Switzerland}

\begin{abstract}
We propose to use the phenomenon of resonant tunneling for the detection of noise.
The main idea of this method relies on the effect of homogeneous broadening of 
the resonant tunneling peak induced by the emission and absorption of collective
charge excitations in the measurement circuit. In thermal equilibrium, the signal 
to noise ratio of the detector as a function of the detector bandwidth (the detector
function) is given by the universal hyperbolic tangent, which is the statement of the 
fluctuation-dissipation theorem. The universality breaks down if non-equilibrium 
processes take place in the measurement circuit. We propose the theory of this phenomenon 
and make predictions for the detector function in case when non-equilibrium noise is 
created by a mesoscopic conductor. We investigate measurement circuit effects and prove
the universality of the classical noise detection. Finally, we evaluate the contribution
of the third cumulant of current and make suggestions of how it can be measured.

\end{abstract}


\maketitle

\section{Introduction} 
\label{intro}

Universalities play an important role in physics, because they point to 
fundamental laws and properties, such as  symmetries, topology, scaling behavior,
and others. Moreover, when broken, they open a door to new physics. Here we wish 
to consider one example that is important in the context of the present work. Recently, 
following the suggestion of Kane and Fisher, \cite{kane-fisher} experiments on 
shot noise in quantum Hall systems\cite{Glattli,dePicciotto} directly measured 
fractional charge of Laughlin quasiparticles. The interpretation of these
experiments invokes a simple argument that weak quasiparticle tunneling is an uncorrelated 
Poisson process which is described by the Schottky formula $S=q\langle I\rangle$,
where $\langle I\rangle$ is the average tunneling current, $S$ is the zero-frequency
noise power of the tunneling current, and $q$ is the fractional charge of quasiparticles.

More rigorously, the Schottky formula follows from the fluctuation
dissipation theorem (FDT), which states that when a tunnel junction weakly connects 
two metallic reservoirs, the following relation generally holds\cite{rogovin-scalapino,footnote1}
\begin{equation}
q\langle I\rangle/S=\tanh(\Delta\mu/2k_BT),
\label{FDT}
\end{equation}
where $\Delta\mu$ is the electro-chemical potential difference applied to the barrier. 
This relation is a generalization of the well-known Callen-Welton FDT which connects the noise power
and the linear response coefficient,
\cite{callen-welton}
and follows from the argument similar to the one used in the linear response 
theory. This implies the universality of the relation (\ref{FDT}), i.e.\ it holds independently of 
the character of the interaction, spectrum of quasiparticles, the geometry 
of a tunnel junction, etc. It is easy to see that the Callen-Welton
theorem and the Schottky formula are the two limits of the relation (\ref{FDT}).

Here we present a simplified derivation of (\ref{FDT}), based on the ``Golden rule''.\cite{LL03} 
Quantum mechanical transition rate between to energy states $E_n$ and $E_m$ is
given by $W_{mn}=2\pi\delta(E_n-E_m)|A_{mn}|^2$, where $A_{mn}\equiv\langle E_m|A|E_n\rangle$
is the matrix element of the tunneling amplitude $A$. Then the average current can be evaluated
as $\langle I\rangle=q\sum_{mn}W_{mn}(\rho_n-\rho_m)$, where 
$\rho_n\equiv\langle E_n|\rho|E_n\rangle$ is the diagonal matrix element of the density operator, 
i.e.\ the probability to find the system in the state $E_n$. When tunneling is weak, forward
and backward tunneling transitions are independent Poisson processes with the dispersions of 
fluctuations equal to mean currents. Therefore, the total noise power is equal to
$S=q^2\sum_{mn}W_{mn}(\rho_n+\rho_m)$.

\begin{figure}[t]
\centerline{\includegraphics[width=6cm]{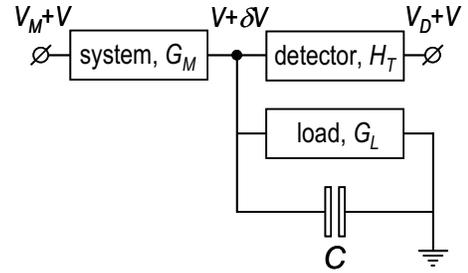}}
\caption{The measurement circuit contains a mesoscopic system which
creates non-equilibrium noise and has the conductance $G_M$, the detector consisting 
of a tunnel junction shunted by the load 
resistor, $G_L$, and the capacitor $C$. The voltage bias $V_M$ and $V_D$ is applied 
to the system and to the tunnel junction, respectively. Fluctuating current through a mesoscopic system
is accumulated on the capacitor and creates the fluctuating potential $\delta V$ across the 
tunneling barrier.}
\vspace{-2mm}
\label{circuit}
\end{figure}

In equilibrium $\rho_{n}\propto\exp(-E_n/k_BT)$, so that $\rho_n=\rho_m$, and the current
vanishes. If the potential difference $\Delta\mu$ is applied between the leads which are
locally at equilibrium, then the density matrix acquires the grand-canonical form 
$\rho(\Delta\mu)=\rho(0)\exp(\Delta\mu N/k_BT)$, where $N$ is number of electrons in one
of the leads. Since the tunneling amplitude changes the number of particles in this lead
by one, then obviously one can write $\rho_m=\exp(\Delta\mu/k_BT)\rho_n$, which immediately 
gives the relation (\ref{FDT}).

From the derivation of the FDT it is obvious that non-equilibrium processes in reservoirs
play a special role, since they may lead to a deviation from the universal relation (\ref{FDT}).
The goal of the present paper is to investigate this phenomenon in the 
case when nonequilibrium processes take place in the electrical (measurement) circuit
to which the tunnel junction is attached. In Fig.\ \ref{circuit} we draw its simplified version 
that contains essential parts: The mesoscopic system which
creates non-equilibrium noise and has the conductance $G_M$, the detector consisting 
of a tunnel junction shunted by the load 
resistor, $G_L$, and the capacitor $C$. One of the important results 
of our paper is that the FDT breaks down in a minimal way, so that some properties 
of the current-to-noise ratio, which contain an important information about non-equilibrium 
processes in the leads, retain their universality. This leads to the idea to use tunnel 
junctions as on-chip detectors of  non-equilibrium noise, which we investigate below in details.

The measurement circuit has been proven to play an important role in the physics of the noise 
detection with the standard measurement technique \cite{cascade-beenakker,cascade-reulet,Reznikov} and 
with the help of on-chip noise detectors.\cite{ESandAJ} It has been established that in the
long time (Markovian) limit the backaction of the measurement circuit on the system leads
to ``cascade corrections'' to statistics of noise.\cite{cascade-nagaev,Geneva_group} 
In order to quantify the circuit effects one solves the Kirhgoff (the current conservation) law 
for the fluctuations of the current through the mesoscopic system $\delta I_M$ and through the load 
resistor $\delta I_D$, and the voltage fluctuations $\delta V$ on the capacitor:
\begin{equation}
\delta V(\omega)=Z(\omega)[\delta I_M(\omega)+\delta I_L(\omega)].
\label{impedance1}
\end{equation}
The circuit impedance is given by 
\begin{equation}
Z(\omega)=R/(1-i\omega\tau_C),
\label{impedance2}
\end{equation}
where $R=1/(G_M+G_L)$ is the differential circuit resistance, and  $\tau_C=RC$ is the circuit 
response time. The Eq.\ (\ref{impedance1}) describes the effect of the system current fluctuations
via the circuit on the tunnel junction which directly detects potential fluctuations. 
The normalized circuit resistance ${\cal R}=G_0R$, where 
$G_0=e^2/2\pi$ is the conductance quantum, plays a role of the dimensionless coupling 
constant, which parametrizes the strength of the circuit effects.
In the present work we assume that coupling is weak, ${\cal R}\ll 1$. 

Quantum noise detectors, the main operating principle of which is based on the resonant tunneling
in a two-level system, where investigated experimentally and theoretically in a number of previous
works. \cite{twolevel-fujisawa,twolevel-aguado,twolevel-schoelkopf,twolevel-onac,twolevel-khrapai,twolevel-gustavsson}
Here we consider two different detectors of this type.
The first one, the double-dot (DD) detector of quantum noise theoretically analyzed by Aguado and Kouwenhoven,  
\cite{twolevel-aguado} is shown in Fig.\ \ref{double-dot}. It consists of two quantum dots, which are strongly coupled to leads
and weakly coupled to each other. To the lowest order in inter-dot coupling the electron 
transport in the DD detector occurs via inelastic transitions between nearest energy
levels of two dots. These transitions are assisted by the emission (absorption) of the
energy $\varepsilon$ to (from) the circuit, where $\varepsilon$ is the inter-dot level 
distance. Away from the resonance the average current through the DD detector is given
by \cite{twolevel-aguado} $\langle I\rangle\sim S_M(\varepsilon)/\varepsilon^2$, where $S_M(\varepsilon)$
is the spectral density of the non-symmetrized correlator of the system current. 
It is easy to see that the parameter $\varepsilon$ plays a role of the bandwidth 
of the detector. 

The operating principle of the second detector, based on the telegraph process 
(TP detector, see Fig.\ \ref{telegraph-process}),  is slightly different. 
It contains two weakly coupled quantum dots, which are electrically isolated
from the circuit, but capacitively coupled to it. Fluctuations of the potential 
on the capacitor, caused by the current noise in the mesoscopic system, lead to 
rare electron transitions between two dots which change the electrical charge 
of, say, the left dot randomly in time. When left dot interacts with a nearby 
quantum point contact (QPC), it randomly switches the QPC current between two 
levels, $I_d$ and $I_u$, leading to the telegraph process. The average 
QPC (detector) current $\langle I_D\rangle$ is a monotonic function of the inter-dot
level distance $\varepsilon$, which changes from one current level to the other,
depending on the average occupation of the left dot. Thus the QPC acts as a
sensitive electrometer of the occupation of the quantum dot, the principle 
demonstrated in early work [\onlinecite{qpc-field}] and 
subsequently elaborated in recent experiments, 
where the real time detection of single-electron tunneling,
\cite{qpc-elzerman,qpc-fujisawa,qpc-schleser,qpc-vandersypen,charge-dicarlo,charge-fujisawa}
the measurement of counting statistics,
\cite{counting-gustavsson1,counting-gustavsson2,counting-gustavsson3}
and the information backaction of a detector\cite{sukhorukov1}
have also been shown.

We denote with $D(\varepsilon,\Delta)$ the current-to-noise ratio for the DD detector and 
call it the detector function. Although $D$ depends on the bias voltage $V_D$,
we choose to represent it as a function of the energy $\Delta=eV_D-\varepsilon$
of the electron-hole pair created in the leads by the elementary tunneling event. 
This energy parametrizes the asymmetry of the detector, because in case $\Delta=0$, or equivalently
$\varepsilon=eV_D$, there is no difference between left and right dot of the DD system. 
Below we prove an important fact that the average current through the QPC of the TP detector, 
after a proper normalization [see Eqs.\ (\ref{ID2}) and (\ref{ID3})], 
is given by the symmetric detector function $D(\varepsilon,0)$. 
When the circuit is at thermal equilibrium, $D(\varepsilon,0)=\tanh(\varepsilon/2k_BT)$ 
according to the FDT. 

The physics of quantum noise detection is quite rich thanks to a number of
energy scales that determine the dynamics of entire system. While these 
energy scales are not important in the case of equilibrium circuit, 
because the FDT holds and leads to the universality, they start to play an 
important role away from equilibrium. First of all, it is an effective 
temperature of the noise source, $\Omega$, which is formally defined by Eq.\
(\ref{energy}). It has a meaning of the energy provided by the system and 
the load. Alternatively, one can think of the correlation time $1/\Omega$ 
of the noise source. Second important parameter is the 
detector bandwidth $\varepsilon$, introduced earlier. Third, the circuit
itself is characterized by the response time $\tau_{C}$ and corresponding
energy scale $1/\tau_C$ (we set $\hbar=1$). Finally, the asymmetry of the DD
detector is characterized by the energy $\Delta$.  

In the weak coupling limit, which we consider throughout the paper, where the 
dimensionless circuit impedance ${\cal R}$ is small,
the detector operates at the {\em Gaussian point}, i.e.\ the contribution
of high order cumulants (irreducible moments of noise) is small. The physical reason 
for this is that in the limit ${\cal R}\ll 1$ the detector only weakly interacts with 
the the noise source, therefore it has to operate for a relatively long time interval
of the order of $1/{\cal R}\Omega$ in order to accumulate a sufficient information about 
the noise. During this time interval many fluctuations contribute to the detector signal, 
so that by virtue of the central limit theorem the resulting noise becomes Gaussian.
In Sec.\ \ref{cumulant} we show how the third cumulant, which is the simplest characteristics 
of the non-Gaussianity, nevertheless can be extracted from the detector output signal. 

The new energy scale $\Gamma_\Omega$ arises in the weak coupling limit due to the 
effect of {\em homogeneous level broadening}: Close to the resonance, 
$\varepsilon\to 0$, the interaction of the detector with the circuit becomes
effectively strong, and inelastic transitions in the detector are assisted by multiple
photon absorption and emission processes. As a result, the detector signal at this point 
acquires a peak as a function of $\varepsilon$ of the width $\Gamma_\Omega\ll\Omega$. The shape
of the peak depends on the circuit details. We distinguish two limiting cases, depending 
on the circuit response time $\tau_C$. In the ``fast'' circuit limit, ${\cal R}\Omega\tau_C\ll 1$,
the peak has a Lorentzian shape and the width $\Gamma_\Omega= 2\pi {\cal R}\Omega$, see Eq.\ (\ref{PE4}). 
In the ``slow'' circuit limit, ${\cal R}\Omega\tau_C\gg 1$, the peak acquires the Gaussian shape
(\ref{PE5}) with the width $\Gamma_\Omega= 2\sqrt{E_C\Omega}$, where $E_C=e^2/C$ is the Coulomb charging
energy of the circuit.

Depending on energy scales following regimes can be distinguished.
In the quantum noise detection regime, $\varepsilon\sim\Omega$, 
the detector signal is due to the inelastic 
tunneling with the absorption or emission of a single  photon of the energy 
$\varepsilon$. The probability of this process is given by Eq.\ (\ref{PE2}).
In case when the circuit is driven away from equilibrium by a coherent 
mesoscopic conductor the symmetric detector function is given by 
Eq.\ (\ref{result1}). For a low circuit impedance, $G_MR\ll 1$,
this expression simplifies and we obtain the result (\ref{result1-dimless}).
The results are summarized in Fig.\ \ref{D}. In the case of strongly asymmetric
detector, $\Delta\gg\Omega$, the detector  function takes the equilibrium 
tangent form (\ref{result2}), which is however shifted by the energy $E_M$
given by Eq.\ (\ref{shift}), which can be viewed as the noise rectification effect.

In the classical noise detection regime, $\varepsilon\ll\Omega$,  
the detector function is simply linear  in $\varepsilon$ [see Eq.\ (\ref{result3})],
with the slope determined by the effective noise temperature $\Omega$.
Thanks to this universality, there is no need to specify the 
mesoscopic system which is measured. Close to the resonance, 
$\varepsilon\sim\Gamma_\Omega$, the inelastic tunneling becomes non-perturbative
despite the small parameter ${\cal R}$, and the $P(E)$ function acquires a peak
of the width $\Gamma_\Omega$. The shape of the peak depends on the circuit 
details, see Eqs.\ (\ref{PE4}) and (\ref{PE5}). Nevertheless, the detector function
retains its universal form (\ref{result3}), so it can be used to extract the noise 
temperature. 

We evaluate the small contribution of the third cumulant of the system current 
in the classical (Markovian) limit and find that it slightly shifts the zero of 
the detector function (\ref{result5}) by the energy $E_3$ which is proportional to
the third cumulant. The coefficient depends on the circuit response time $\tau_C$
and is evaluated in the case of fast and slow circuit, see Eqs.\ (\ref{result6})
and (\ref{result7}). The  total third cumulant of the system current contains cascade 
corrections, which depend on the circuit response time. In the case of fast circuit 
the cascade corrections are given by Eq.\ (\ref{cumulant1}), i.e. they are   
those introduced by Nagaev in Ref.\ [\onlinecite{cascade-nagaev}]. In the slow circuit 
case the detector measures equal time fluctuations of the potential on the capacitor, 
and the cascade corrections in this case are given by Eq.\ (\ref{cumulant2}), 
as predicted in Ref.\ [\onlinecite{ESandAJ}] and measured in Ref.\ [\onlinecite{Reznikov}].
We finally note, that the third cumulant of current may be extracted from the 
shift of the detector function using the technique recently introduced in 
experiments on the mesoscopic threshold detectors.\cite{threshold-pekola,threshold-huarg} 
The universality of the detector function in the classical noise detection regime,
proven in Sec.\ \ref{universalities}, may become crucial for the success of this procedure.

The rest of the paper is organized as follows.  After reviewing the $P(E)$-theory of 
tunneling in Sec.\ \ref{PE}, we focus on the Gaussian noise case in Sec.\ \ref{gaussian}
and classify the measurement circuit effects according to the circuit response time. 
In Sec.\ \ref{detectors} we analyze quantum noise detectors based on the resonant 
tunneling effect and connect the detector function $D(\varepsilon,\Delta)$
to the current-to-noise relation for tunnel junctions. We use results of the $P(E)$-theory 
in Sec.\ \ref{FDR} to calculate the detector function in quantum and classical noise 
detection regimes. In Sec.\ \ref{universalities} we prove that the detector function
is universal in the classical noise detection regime, i.e.\ it is independent of
the measurement circuit details. Finally, in Sec.\ \ref{cumulant} we investigate 
the third cumulant contribution to the detector function including the circuit 
cascade corrections. The section \ref{conclusion} outlines further
directions of research.

\section{Reminder on $P(E)$-theory of tunneling} 
\label{PE}

The purpose of this section is to remind essential steps of the $P(E)$-theory of
photon-assisted tunneling.\cite{ingold-nazarov} In additional, we extend the theory in order to take 
into account weak non-Gaussian 
effects in noise. The tunnel junction, attached to two metallic leads, 
is described by the Hamiltonian:
\begin{equation}
H=\sum_k\varepsilon_k (c^{\dagger}_kc_k+d^{\dagger}_kd_k)+ H_T,
\label{leads}
\end{equation}  
where $c_k$ and $d_k$ are the electron operators in the left and right lead,
respectively, and $H_T$ is the tunneling Hamiltonian.
It can be written as \cite{ingold-nazarov}
\begin{equation}
H_T=A+A^\dagger, \quad A=e^{i\phi}\sum_{pk}T_{pk}d^{\dagger}_pc_k,
\label{tunneling}
\end{equation}
where the amplitude $A$ transfers the electron from left to right, and 
the phase factor $e^{i\phi}$ changes the charge on the capacitor by $-e$. The last 
fact follows from the charge quantization $[\phi,Q]=ei$. Then the charge Hamiltonian 
$H_C=Q^2/2C$ generates the equation of motion for the phase operator:
$\dot\phi=e\delta V$. We thus assume that the interaction of electrons with 
the collective charge excitations in the electrical circuit is generated solely 
via tunneling. 

Next we evaluate the average tunneling current $\langle I_D \rangle$ and the zero-frequency  
noise power $S_D=\int dt \langle\delta I_D(t)\delta I_D(0)\rangle$. We define the tunneling
current operator as $I_D\equiv edN_L/dt=ie(A-A^\dagger)$,
where $N_L=\sum_kc^\dagger_kc_k$ is the number of electrons in the left lead.
To leading order in the tunneling Hamiltonian (\ref{tunneling}), we can write
\begin{subequations}
\begin{eqnarray}
\langle I_D\rangle &=& e\int dt\langle[A(t),A^\dagger(0)]\rangle,
\label{current1}\\
S_D &=& e^2\int dt \langle\{A(t),A^\dagger(0)\}\rangle.
\label{noise1}
\end{eqnarray} 
\label{detector1}
\end{subequations}
Substituting $A$ from Eq.\ (\ref{tunneling}) to Eqs.\ (\ref{detector1}), and
tracing out electronic operators, we finally obtain:
\begin{widetext}
\begin{subequations}
\begin{eqnarray}
\langle I_D\rangle &=& 2\pi e\int\!\!\!\int dEdE'\nu_R(E)\nu_L(E') 
\{P_{LR}(E-E'+eV_D)f(1-f')-P_{RL}(E'-E-eV_D)f'(1-f)\},
\label{current2}\\
S_D &=& 2\pi e^2\int\!\!\!\int dEdE'\nu_R(E)\nu_L(E') 
\{P_{RL}(E-E'+eV_D)f(1-f')+P_{LR}(E'-E-eV_D)f'(1-f)\},
\label{noise2}
\end{eqnarray} 
\label{detector2}
\end{subequations}
\end{widetext}
where $f=f_F(E)$ and $f'=f_F(E')$ are the equilibrium distributions in the leads, and 
$\nu_L$ and $\nu_R$ are the electronic densities of states.   
Here $P_{LR}(E)$ and $P_{RL}(E)$ are the probability distributions of the emission 
(absorption) of a 
collective charge excitation of energy $E$, caused by inelastic tunneling of an electron 
from right to left lead and vice versa:   
\begin{subequations}
\begin{eqnarray}
P_{LR}(E) &=& \frac{1}{2\pi}\int \, dt\, 
e^{iEt}\langle e^{i\phi(t)}e^{-i\phi(0)}\rangle, 
\label{LR}\\
P_{RL}(E) &=& \frac{1}{2\pi}\int \, dt\, e^{iEt}\langle e^{-i\phi(t)}e^{i\phi(0)}\rangle.
\label{RL}
\end{eqnarray}
\label{PE1}
\end{subequations}

In general, the phase correlation functions in (\ref{PE1}) can be expanded 
in terms of the noise cumulants. However, every cumulant comes with extra
coupling constant ${\cal R}\ll 1$. Therefore, we keep only first two 
nonvanishing cumulants
\begin{eqnarray}
J_2(t)&=&\frac{1}{2}\langle\phi^2(t)-2\phi(t)\phi(0)+\phi^2(0)\rangle
\label{J0-2}\\
J_3(t)&=&\frac{1}{6}\langle\phi^3(t)-3\phi^2(t)\phi(0)\nonumber\\
&&\qquad+3\phi(t)\phi^2(0)-\phi^3(0)\rangle
\label{J0-3}
\end{eqnarray}
and write
\begin{subequations}
\begin{eqnarray}
P_{LR}(E) &=& \frac{1}{2\pi}\int \, dt\, e^{iEt-J_2(t)-iJ_3(t)}, 
\label{LR-expansion}\\
P_{RL}(E) &=& \frac{1}{2\pi}\int \, dt\, e^{iEt-J_2(t)+iJ_3(t)}. 
\label{RL-expansion}
\end{eqnarray}
\label{PE-expansion}
\end{subequations}
We postpone the discussion of the third cumulant effect till Sec.\ \ref{cumulant} 
and for a moment assume that the noise is Gaussian.

\section{Gaussian noise}
\label{gaussian} 

We now set $J_3=0$ and write $P_{LR}=P_{RL}\equiv P$, where 
\begin{equation}
P(E) = \frac{1}{2\pi}\int \, dt\, e^{iEt-J_2(t)},
\label{PE-Gaussian}
\end{equation}
Note that in Eq.\ (\ref{J0-2}) each term of the form $\langle\phi^2\rangle$ contain a classical 
contribution, which in the long time limit is proportional to time.\cite{footnote2}  
This is a consequence of the 
Brownian motion of the phase ``pushed'' by a fluctuating potential.  
However, these potentially dangerous 
terms cancel, and Eq.\ (\ref{J0-2}) can be rewritten in the form 
\begin{equation}
J_2=\frac{1}{2}\langle(\Delta\phi)^2\rangle+\frac{1}{2}\langle[\phi,\Delta\phi]\rangle,\quad
\Delta\phi\equiv \phi(t)-\phi(0),
\label{proper-J}
\end{equation}
so that it does not contain divergences. In this equation the first term can be interpreted
as a classical contribution, which is proportional to time in the long time limit, and the 
second term is pure quantum.
Using Eq.\ (\ref{impedance1}), we obtain
\begin{equation} 
J_2(t)=G_0\int  \frac{d\omega\, S(\omega)}{\omega^2+\eta^2}|Z(\omega)|^2(1-e^{-i\omega t}),\quad
\eta\to 0
\label{J1}
\end{equation}
where $S(\omega)=S_M(\omega)+S_L(\omega)$ is the power of the total noise created in the circuit,
and $S_M$ and $S_L$ are the non-symmetrized correlators of the mesoscopic system and of the load 
resistor
\begin{subequations}
\begin{eqnarray}
S_M(\omega) &=& \int dt e^{i\omega t}\langle\delta I_M(t)\delta I_M(0)\rangle,
\label{SM}\\
S_L(\omega) &=& \int dt e^{i\omega t}\langle\delta I_L(t)\delta I_L(0)\rangle,
\label{SL}
\end{eqnarray} 
\label{noises1}
\end{subequations}

Next we note that in the weak interaction case ${\cal R}=G_0R\ll 1$ considered here,
$J_2(t)$ is usually small. For instance, in equilibrium $RS(\omega)=2k_BT$, so that for
$t\sim 1/k_BT$ the correlator given by Eq.\ (\ref{J1}) can be roughly estimated as $J_2\sim {\cal R}$.
Therefore, we expand the exponential on the right hand side of Eq.\ (\ref{PE-Gaussian})
and obtain:
\begin{equation}
P(E)=P_0\delta(E)+G_0|Z(E)|^2S(E)/E^2,
\label{PE2}
\end{equation}
where $P_0$ is the probability of the elastic process, fixed by the normalization $\int dEP(E)=1$.
The probability of the inelastic process is proportional to the non-symmetrized correlator
$S(E)$, \cite{twolevel-aguado,asymmetry-lesovik,asymmetry-gavish1,asymmetry-gavish2} 
and at relatively large energies is sensitive 
to quantum fluctuations.

Special care however has to be taken about the long time limit in Eq.\ (\ref{J1}),
since growing with time classical contribution to $J_2$ may compensate smallness of ${\cal R}$.
The Fourier integral cuts off a small region around $\omega=0$, where the noise is
classical, and the noise power can be approximately replaced with $S(0)$. The important note
is in order: Quantum effects, which lead to the interaction induced suppression of tunneling, 
i.e.\ to so called dynamical Coulomb blockade effect, \cite{ingold-nazarov,blockade-altimiras}
are not neglected. They are fully taken into account in Eq.\ (\ref{PE2}), and subsequently,
in Sec.\ \ref{FDR}. However, at the energy scale of interest here their contribution to the 
long time asymptotic is small.  
We now focus on the long time limit and consider the cases of fast and slow circuit, 
depending on the circuit response time $\tau_C$.

\subsection{Fast circuit}
\label{fast}

We first assume that the relevant time scale is longer than $\tau_C$, and therefore set $Z(\omega)=R$.
From Eq.\ (\ref{J1}) we find:
\begin{equation} 
J_2(t)=2\pi {\cal R}\Omega\,\{|t|+i\partial_\omega S(0)/S(0)\,{\rm sign}(t)\},
\label{J2}
\end{equation}
where the energy scale $\Omega$ is the circuit noise temperature:
\begin{equation}
\Omega\equiv (1/2) R S(0),
\label{energy}
\end{equation}
Note that although the interaction is weak, ${\cal R}\ll 1$, in the long time limit $|t|\sim 1/{\cal R}\Omega$
the exponential in Eq.\ (\ref{PE1}) cannot be expanded. We then use the result (\ref{J2}) and obtain:
\begin{equation}
P(E)=\frac{{2\cal R}\Omega}{E^2+(2\pi {\cal R}\Omega)^2}\;[1+E\partial_\omega S(0)/S(0)],
\label{PE3}
\end{equation}
which is consistent with the result (\ref{PE2}) in the limit $E\gg{\cal R}\Omega$.

Thus we find that in the limit, $|t|\Omega\sim 1/{\cal R}\gg 1$, the multiple photon processes 
lead to the broadening of the $\delta$-function in Eq.\ (\ref{PE2}), so that it is replaced 
with the Lorentzian peak. One can now use Eq.\ (\ref{impedance2}) to check that the assumption 
$Z(\omega)=R$ is justified if ${\cal R}\Omega\tau_C\ll 1$. This means that the response of the 
circuit to current fluctuations is instantaneous, and the phase fluctuations are Markovian on the 
time scale of interest. 

The asymmetry of $P(E)$ given by the second term in Eq.\ (\ref{PE3}) is weak: $E\partial_\omega S/S\sim{\cal R}$. 
Interestingly, the expression $\partial_\omega S(0)=\int dt(it/2)\langle[I(t),I(0)]\rangle $ 
coincides with the Kubo formula for the differential conductance, $1/R=\partial_V\langle I\rangle$. 
Therefore we obtain $\partial_\omega S(0)=1/R$, and alternatively,
\begin{equation}
\frac{\partial_\omega S(0)}{S(0)}=\frac{1}{2\Omega}\,.
\label{identity}
\end{equation} 
Thus  one can express 
the asymmetry in (\ref{PE3}) in terms of the noise temperature alone:
\begin{equation}
P(E)=\frac{{\cal R}(2\Omega+E)}{E^2+(2\pi {\cal R}\Omega)^2},
\quad {\cal R}\Omega\tau_C\ll 1.
\label{PE4}
\end{equation}

\subsection{Slow circuit}
\label{slow}

Next we consider the opposite limit, ${\cal R}\Omega\tau_C\gg 1$, when circuit responds
slowly to current fluctuations. In this case it is the singularity
in $Z(\omega)$ cuts off the integral in (\ref{J1}) at small frequencies $\omega\sim 1/\tau_C$.
Using the impedance (\ref{impedance2}) and the relations (\ref{energy}) and (\ref{identity}), 
we obtain:
\begin{equation}
J_2(t)=\pi{\cal R}(\Omega/\tau_C)t^2+i\pi{\cal R}(1/\tau_C)t
\label{J3}
\end{equation} 
The first term in this equation has a simple interpretation. We note that it can be also 
obtained by considering the phase $\phi$ as classical variable, and writing 
$\phi(t)-\phi(0)=e\delta Vt$, because the variation of the potential is slow. 
Then, the equation (\ref{proper-J}) leads to 
$J_2(t)=(e^2/2)\langle(\delta V)^2\rangle t^2$, which together with Eq.\
(\ref{impedance2}) gives the first term in (\ref{J3}). Thus the phase correlator is 
determined by the equal time correlator of the potential.
We will rely on this interpretation in Sec.\ \ref{cumulant}. 

The second term in the equation (\ref{J3}) has a quantum nature. It slightly shifts 
the energy in $P(E)$ given by Eq.\ (\ref{PE-Gaussian})
and leads to the asymmetry of the distribution.
This term is small, therefore the Fourier transform in
(\ref{PE1}) can be written as
\begin{equation}
P(E)=\frac{1+E/2\Omega}{\sqrt{4\pi E_C\Omega}}\exp\left(-\frac{E^2}{4E_C\Omega}\right),
\quad {\cal R}\Omega\tau_C\gg 1,
\label{PE5}
\end{equation} 
where $E_C=e^2/2C$. Thus we see that the dissipative properties of the circuit, determined
by the resistance $R$, do not enter the final result. This is related to the fact,
that slow fluctuations of charge on the capacitor obey the low of 
the equipartition of energy, $(1/2C)\langle(\delta Q)^2\rangle=\Omega/2$, as it
were in equilibrium. 

We remark that in the intermediate regime ${\cal R}\Omega\tau_C\sim 1$ the exact shape 
of the zero energy peak in $P(E)$ is more complex and depends on details of the circuit. 
Nevertheless, as we show in Sec.\ \ref{universalities}, the fluctuation-dissipation 
relations remain insensitive to these details. Finally, we also note that these our
results on the asymmetry in $P(E)$ were published in Ref.\ [\onlinecite{jonathan}]. 
Recently the asymmetry was found in the experiment [\onlinecite{harris}]
and theoretically discussed in Ref.\  [\onlinecite{averin}].

\begin{figure}[t]
\centerline{\includegraphics[width=5cm]{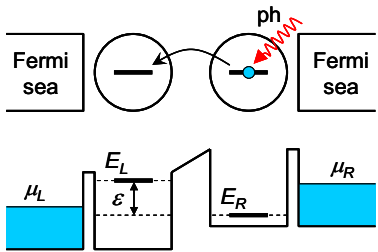}}
\caption{The DD detector operates as shown on the upper panel. 
The absorption of the quantum of the collective charge excitation
of the circuit leads to the inelastic electron transition between two 
weakly connected quantum dots. Because dots are strongly connected 
to two metallic reservoirs, multiple random transitions generate 
current through the detector and the current noise. Lower panel 
show the energy diagram of the detector and most important parameters.}
\vspace{-2mm}
\label{double-dot}
\end{figure} 

\section{Quantum noise detectors} 
\label{detectors}

We have briefly discussed two types of quantum noise detectors in the introduction.
Here we analyze them in details and show that their properties are determined
by the $P(E)$ function, obtained in previous section. Starting with the DD detector,
we first assume that tunneling between two dots is a weakest process. In this simple 
case the transport can be described by lowest order in tunneling, so that the result
(\ref{detector2}) of previous section fully applies. Moreover, a weak coupling 
of the dots to the reservoirs leads to the broadening of the dot levels, so that 
the densities of states acquire a Breit-Wigner form 
$\nu_{\alpha}=(\Gamma_\alpha/\pi)/[(E-E_\alpha)^2+\Gamma_\alpha^2]$, $\alpha=L,R$. 

If the noise temperature is small, so that  $\Gamma_\Omega<\Gamma_\alpha$, 
then the elastic
transport dominates the photon-assisted inelastic transitions. In this case
the left and right lead are approximately at thermal equilibrium, and the FDT holds. 
The most efficient noise detection takes place for a relatively strong noise 
in the circuit, $\Gamma_\Omega>\Gamma_\alpha$, when the homogeneous level broadening
dominates the quantum effect. In this case Breit-Wigner resonances
can be replaced by delta functions, $\nu_{\alpha}=\delta(E-E_\alpha)$, $\alpha=L,R$, 
where $E_L$ and $E_R$ are the energies of dot levels counted from the local Fermi
level in the left and right lead, see Fig.\ \ref{double-dot}. 
Substituting delta functions to Eqs.\ (\ref{detector2}) and using 
$f'(1-f)=f(1-f')e^{(E_R-E_L)/k_BT}$ for the current
to noise ratio we obtain the following function:
\begin{equation}
\frac{e\langle I_D\rangle}{S_D}\equiv D(\varepsilon, \Delta)=
\frac{P(\varepsilon)e^{\Delta/k_BT}-P(-\varepsilon)}
{P(\varepsilon)e^{\Delta/k_BT}+P(-\varepsilon)}.
\label{FDT1}
\end{equation}
Here the tunable level distance $\varepsilon\equiv E_R-E_L+eV_D$ is the detector bandwidth, 
and the energy of the electron-hole pair $\Delta\equiv E_L-E_R$ parametrizes the asymmetry 
of the detector. The detector function $D(\varepsilon, \Delta)$ will be analyzed in details 
in the next section. Below we show that the properties of the TP detector are determined 
by the symmetric variant of this function, $D(\varepsilon, 0)$.

\begin{figure}[t]
\centerline{\includegraphics[width=7cm]{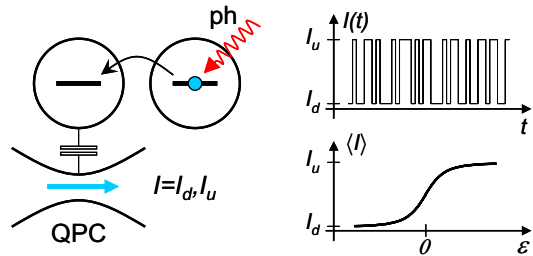}}
\caption{The TP detector consist of a double-dot system, which is capacitively 
coupled to a QPC. When one of the dots is charged, it pinches off the QPC and 
thus changes the current through it from the upper level $I_u$ to the lower level
$I_d$. Random interdot transitions, caused by the emission and absorption of the
collective charge excitations of the circuit, lead to random switching of the
QPC current. The resulting telegraph process is shown on the upper right panel.
The average current through the QPC, shown on the lower right panel, develops a
smooth crossover between two current levels as a function of the DD level
distance $\varepsilon$.}
\vspace{-2mm}
\label{telegraph-process}
\end{figure}

We evaluate the average current through the QPC capacitively coupled to the DD, see
Fig.\ \ref{telegraph-process}.
Switching of the DD from one state to another changes the current $I_D$ through
the QPC from the low level $I_d$ to the high level $I_u$. The probability
to find the DD in the lower and upper state are given by $P_d=\gamma_d/(\gamma_u+\gamma_d)$
and $P_u=\gamma_u/(\gamma_u+\gamma_d)$, where $\gamma_d$ and $\gamma_u$
are the switching rates. Then the average current is given by 
$\langle I_D\rangle=I_uP_u+I_dP_d$ and is equal to
\begin{equation}
\langle I_D\rangle=\frac{I_u\gamma_u+I_d\gamma_d}{\gamma_u+\gamma_d}.
\label{ID1}
\end{equation} 

It is convenient to rewrite the detector current  $\langle I_D\rangle$ in the dimensionless
form:
\begin{equation}
{\cal I}_D\equiv\frac{2 \langle I_D\rangle -(I_u+I_d)}{I_u-I_d}=\frac{\gamma_u-\gamma_d}{\gamma_u+\gamma_d},
\label{ID2}
\end{equation} 
so that it acquires the maximum value ${\cal I}_D=1$ when $\gamma_u\gg\gamma_d$ and the upper level
is occupied, $P_u=1$, and ${\cal I}_D=-1$ in the opposite case, when mostly the lower level is occupied.

Next we assume that one of the dots is strongly coupled to the circuit capacitor.
Then with a good approximation switching of the DD changes the charge of the capacitor 
by the value $e$, so that the interdot coupling is proportional to $e^{i\phi}$.\cite{footnote3} 
Assuming the interdot coupling is weak compared to the width of levels, 
one can evaluate the switching rate using the Golden rule 
approximation with the result $\gamma_u\propto P(\varepsilon)$ and $\gamma_d\propto P(-\varepsilon)$, 
where $\varepsilon$ is the DD level distance. Therefore, using the result (\ref{ID2}) we obtain:
\begin{equation}
{\cal I}_D(\varepsilon)= \frac{P(\varepsilon)-P(-\varepsilon)}{P(\varepsilon)+P(-\varepsilon)}=
D(\varepsilon, 0),
\label{ID3}
\end{equation} 
i.e.\ the normalized average current through the QPC as a function of the tunable level distance
is given by the symmetric variant of the detector function. 

\section{Fluctuation-dissipation relations} 
\label{FDR}

In this section we investigate how non-equilibrium processes in the circuit lead to a 
breakdown of the FDT. We first focus on
the inelastic regime $\varepsilon > {\cal R}\Omega$, where Eq.\ (\ref{PE2}) applies,
and later consider the classical regime described by Eqs.\ (\ref{PE4}) and (\ref{PE5}). 
Substituting Eq.\ (\ref{PE2}) to the definition (\ref{FDT1}), we obtain
\begin{equation}
D(\varepsilon, \Delta)=
\frac{S(\varepsilon)e^{\Delta/k_BT}-S(-\varepsilon)}
{S(\varepsilon)e^{\Delta/k_BT}+S(-\varepsilon)},
\label{FDT2}
\end{equation}
where, we remind, $\varepsilon=eV_D-\Delta$ is the interdot level distance.
Thus all the circuit details cancel from the final result, and the exact form of the 
function $D$ is determined solely by a non-symmetrized 
correlator of the current fluctuations in the circuit. 

In order to make further progress, we have to specify the model of the current source. 
The load resistor may be considered as a macroscopic system that creates an equilibrium
current noise. Non-equilibrium processes are generated by the mesoscopic system alone.
An interesting and experimentally important example of the mesoscopic system is a 
coherent mesoscopic conductor, which is fully characterized by a set of transmission 
eigenvalues, $T_n$, $n=1,\ldots,N$. Using the scattering theory,\cite{blanter-buttiker} 
one obtains the following expression for the non-symmetrized 
current correlator
\begin{eqnarray}
S_M(\omega)&=&G_0\sum_n \bigg\{2T_n^2\,F(\omega)+T_n(1-T_n)\nonumber\\
&\times& [F(\omega+eV_M)+F(\omega-eV_M)]\bigg\},
\label{SM1}
\end{eqnarray}
where 
\begin{equation}
F(\omega)\equiv \frac{\omega}{1-e^{-\omega/k_BT}},
\label{F}
\end{equation}
and we assumed that transmission eigenvalues $T_n$ are energy independent.
Using $F(\omega)-F(-\omega)=2\omega$ we now check that 
indeed $S_M(\omega)-S_M(-\omega)=2\omega G_M$, where the conductance $G_M=G_0\sum_nT_n$.
The same relation obviously holds for the macroscopic resistor.

In equilibrium Eq.\ (\ref{SM1}) gives $S=S_M+S_L=2(G_M+G_L)F(\omega)$, 
a well known result for the non-symmetrized noise power. It satisfies the detailed
balance relation $S(-\omega)=e^{-\omega/k_BT}S(\omega)$. Substituting this relation to  
Eq.\ (\ref{FDT2}), we arrive at the equilibrium function $D=\tanh(eV_D/2k_BT)$,
in agreement with the FDT. 
If the load conductance is large, $G_L\gg G_M$, the equilibrium noise of the load 
resistor may dominate in the circuit noise. In this case the function $D$ may retain 
its equilibrium form,  even if the mesoscopic conductor is biased. It is therefore 
interesting to consider the strong bias regime, $eV_M>k_BT/R$, so that the equilibrium 
noise contribution can be neglected. Three important cases, which deserve special consideration,
are discussed below.

\subsection{Symmetric detector, $\Delta=0$}
\label{symmetric} 

This case is most relevant for the TP detector, which is symmetric detector. 
Using the zero-temperature limit
$F(\omega)=\omega$ in Eq.\ (\ref{SM1}) and substituting the result to the Eq.\ (\ref{FDT2}),
we obtain an important result
\begin{equation}
D(\varepsilon,0)=
\frac{\varepsilon}{{\cal F}G_MR\left(eV_M-|\varepsilon|\right)+|\varepsilon|},\quad  
\mbox{for}\quad |\varepsilon|<eV_M,
\label{result1}
\end{equation}
and $D=\pm 1$, otherwise. Here ${\cal F}\equiv\sum_nT_n(1-T_n)/\sum_nT_n$ is the
Fano factor of the system noise. Note that the slope of $D$ at $\varepsilon=0$
is equal to $1/(e{\cal F}G_MRV_M)=1/(2\Omega)$, where $\Omega$, we remind, is the circuit 
noise temperature. Interestingly, as we show below, this slope is universal 
and same for an arbitrary mesoscopic conductor and arbitrary circuit.

\begin{figure}[t]
\centerline{\includegraphics[width=7.5cm]{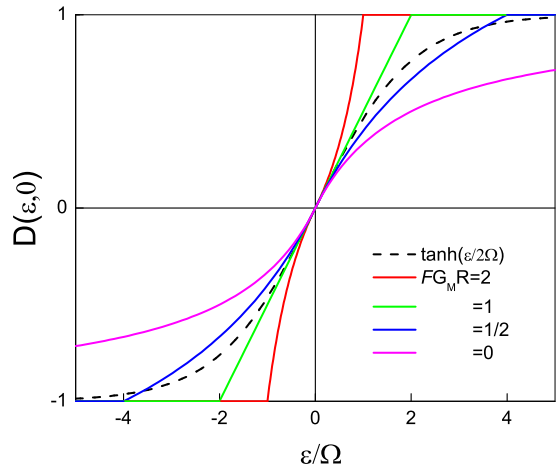}}
\caption{The symmetric detector function $D(\varepsilon,0)$ is plotted versus 
normalized level spacing $\varepsilon/\Omega$ for different values of the parameter 
${\cal F}G_MR$. Typically, $D$ is concave function of $\varepsilon$, although for 
a super-Poissonian noise ${\cal F}>1$ it may become convex. Note that in the 
limit ${\cal F}G_MR\ll 1$ the detector function has a power law behavior 
as compared to the exponential behavior of the equilibrium $D=\tanh(\varepsilon/2\Omega)$
shown by the dashed line.}
\vspace{-2mm}
\label{D}
\end{figure}

In the case of a very low load impedance, $G_MR\ll 1$, the result (\ref{result1})
can be written in the dimensionless form
\begin{equation}
D(\varepsilon,0)=\frac{\varepsilon/\Omega}{2+|\varepsilon/\Omega|},
\label{result1-dimless}
\end{equation}
It is plotted on Fig.\ \ref{D} together with the equilibrium $D(\varepsilon,0)=
\tanh(\varepsilon/2\Omega)$. 
The non-equilibrium $D$ has a power-low asymptotic
at $|\varepsilon|\to \infty$, while the equilibrium one shows an 
exponential behavior. Note also that for ${\cal F}G_MR>1$ the 
detector function $D(\varepsilon,0)$ is convex, which could be 
considered a signature of a super-Poissonian noise.  

\subsection{Asymmetric detector}
\label{asymmetric} 

We consider a circuit far away from equilibrium, $eV_M>k_BT/R$. We expect that 
it might be difficult to adjust the DD detector precisely to cancel the asymmetry,
therefore we first assume that the asymmetry is strong, $|\Delta|\gg eV_M$.  
Then, looking at the result (\ref{FDT2}), we expect that $D=1$, i.e.\ the noise
of the DD detector is Poissonian. In fact, more careful analysis shows that the 
strong asymmetry simply shifts the zero of the function $D$. This is because 
small or large value of the exponential $e^{\Delta/k_BT}$ may be compensated 
by the opposite effect in $S$ due to the activation processes. 

Looking at the result (\ref{SM1}) and (\ref{F}) 
we find that for positive $\omega$ the dominant contribution is
$S_M(-\omega)={\cal F}G_M(\omega-eV_M)e^{(eV_M-\omega)/k_BT}$ 
due to such activation processes. The contribution of the load resistor, $S_L(-\omega)$ 
is small. On the other hand, both the load resistor and the mesoscopic system
contribute to the term in Eq.\ (\ref{FDT2}): $S(\omega)=2\omega/R$.
Neglecting terms $eV_M$ and $eV_D$ compared to $\Delta$, we finally obtain:
\begin{equation}
D(\varepsilon, \Delta)=\tanh\left[\frac{eV_D+{\rm sgn}(\Delta)E_M}{2k_BT}\right],
\label{result2} 
\end{equation}
where the energy shift $E_M$ is given by
\begin{equation}
E_M=eV_M+k_BT\log({\cal F}G_MR/2).
\label{shift} 
\end{equation}
Thus we arrive at the remarkable result that the only role of $\Delta$ is to fix
the sign of the energy shift in (\ref{result2}). This fact is easily understood 
when we notice that the energy shift can be viewed as a {\em drag} or noise 
{\em rectification} effect, the direction of which depends on the sign of $\Delta$. 

Interestingly, there is an additional contribution to $E_M$ in form of the logarithm,
which contains the system Fano factor. It is exactly same parameter that also 
appears in the symmetric case (\ref{result1}). This additional shift may be 
interpreted as originating from high energy excitations that create the shot
noise in mesoscopic system. Its explicit form depends on the assumption we made
that the system is a coherent mesoscopic conductor. Therefore, it would be 
interesting to consider other examples of mesoscopic systems which may change 
the result (\ref{result2}) and (\ref{shift}). 

\subsection{Classical noise regime}
\label{classical} 

So far we have discussed essentially quantum regime of the noise detection,
where specific form of the function $D$ depends on the choice 
of the system. In the rest of the paper we concentrate on the 
classical Markovian limit, which corresponds to a small detector 
bandwidth $\varepsilon$, and demonstrate a number of universalities. 

We note that although the function $D$ in Eqs.\ (\ref{result1}) and (\ref{result1-dimless})
behaves regularly at $\varepsilon=0$, it has been obtained in the limit 
$\varepsilon > {\cal R}\Omega$ using the result (\ref{PE2}). If the detector bandwidth $\varepsilon$ tends
to zero, $P(\varepsilon)$ given by Eq.\ (\ref{PE2}) diverges and has to be replaced
with the resummed version (\ref{PE4}). 
Natural question arises is whether a considerable change in $P(\varepsilon)$,
including appearance of the peak at $\varepsilon=0$,
affects the symmetric detector function (\ref{result1}). The answer is no. 
We first check this for a fast and slow circuit limit, and prove the universality
in the next section.

Indeed, substituting either the function (\ref{PE4}) or the function
(\ref{PE5}) into Eq.\ (\ref{FDT1}), for $\Delta=0$
we obtain
\begin{equation}
D(\varepsilon,0)=\frac{P(\varepsilon)-P(-\varepsilon)}{P(\varepsilon)+P(-\varepsilon)}
=\frac{\varepsilon}{2\Omega}, \qquad \mbox{if $\varepsilon\ll\Omega$},  
\label{result3}
\end{equation}
where, we remind, $\Omega$ is the circuit noise temperature (\ref{energy}).
This result agrees with (\ref{result1}) as $\varepsilon\to 0$. We stress
however, that the result (\ref{result3}) is more general, since its derivation
does not relay on the scattering theory \cite{blanter-buttiker} for a mesoscopic 
coherent conductor. 

We now in the position to investigate the effect of asymmetry. Restricting ourselves
to the classical regime, $\varepsilon\ll \Omega$, we write that generally
$P(\varepsilon)=P_0(\varepsilon)(1+\varepsilon/2\Omega)$, where 
$P_0(\varepsilon)=P_0(-\varepsilon)$ is the classical contribution. Substituting 
this expression to the Eq.\ (\ref{FDT1}), we arrive at 
\begin{equation}
D(\varepsilon,\Delta)=\tanh(\delta)+\frac{\varepsilon}{2\Omega\cosh^2(\delta)}, 
\qquad \delta=\frac{\Delta}{2k_BT}.  
\label{result3-g}
\end{equation} 
Note that this result does not contradict the strongly asymmetric case (\ref{result2}),
because here we assume $\varepsilon\ll \Omega$. It implies that at small asymmetry the 
zero of the detector function is shifted:
\begin{equation}
D(\varepsilon,\Delta)=\frac{\varepsilon+E_2}{2\Omega},\qquad
E_2=(\Omega/k_BT)\Delta.
\label{result3-sh}
\end{equation}
Again, this shift is solely due to the second cumulant of current noise,
and can be interpreted as an asymmetry induced noise rectification effect. 
This fact is important for the discussion in Sec.\ \ref{cumulant}.

\section{Universality of classical noise detection regime} 
\label{universalities}

In order to arrive at the result (\ref{result3}), we used Eqs.\ (\ref{PE4}) and (\ref{PE5})
for a fast and slow circuit, respectively. In general, the shape of the peak 
in $P(E)$ depends on the circuit response time $\tau_C$ and circuit details via the 
impedance function $Z(\omega)$. In this section we show that, surprisingly, in the 
classical limit $\varepsilon\ll\Omega$ the detector function retains its form
(\ref{result3}) and (\ref{result3-g}).

We return now to the equations (\ref{PE-Gaussian}) and (\ref{J1}), and assume that the 
circuit impedance $Z(\omega)$ is arbitrary. The only requirement that we impose 
is that the interaction is weak: ${\cal R}=G_0Z(0)\ll 1$. We focus on the long 
time limit of $J_2(t)$, so that the integral in (\ref{J1}) comes
from small frequencies $\omega\sim {\cal R}\Omega$, where the noise power can 
be approximately expanded as $S(\omega)=S(0)+\partial_{\omega}S(0)\,\omega$.
Consequently, $J_2(t)$ acquires two contributions that can be written as
\begin{equation}
J_2(t)=G_0S(0)H(t)+iG_0\partial_\omega S(0)\partial_tH(t),
\label{J4}
\end{equation}
where 
\begin{equation}
H(t)=\int\frac{d\omega}{\omega^2}|Z(\omega)|^2[1-\cos(\omega t)].
\label{H}
\end{equation}

We are interested in time scales $t\sim 1/({\cal R}\Omega)$,
where the first term in (\ref{J4}) is of order 1, and the peak of
the function $P(E)$ is formed. Then the second term in (\ref{J4})
is of the order of ${\cal R}$, i.e.\ it is always small. Therefore,
its contribution to the exponential in Eq.\ (\ref{PE-Gaussian})
should be expanded, giving the odd part of the $P(E)$ function. 
Thus we obtain 
the following result:
\begin{subequations}
\begin{eqnarray}
P(E)+P(-E)&=&\pi^{-1}\!\!\int dt\exp[-G_0S(0)H(t)]\nonumber\\
&\times&\cos(Et),
\label{PE6}\\
P(E)-P(-E)&=&\pi^{-1}\!\!\int dt\exp[-G_0S(0)H(t)]\nonumber\\
&\times&G_0\partial_\omega S(0)\partial_tH(t)\sin(Et).
\label{PE7}
\end{eqnarray}
\label{PE-general}
\end{subequations}

It is easy to see that by the integration by parts the Eq.\ (\ref{PE7}) 
can be presented in the same form as Eq.\ (\ref{PE6}). Thereby, independently
of the exact function $Z(\omega)$, we 
arrive at the most general result for the classical noise regime, 
$E\ll\Omega$:
\begin{equation}
\frac{P(E)-P(-E)}{P(E)+P(-E)}=\frac{\partial_\omega S(0)}{S(0)}\,E.
\label{result4} 
\end{equation}
Using again the result (\ref{identity}), we arrive at the Eq.\ (\ref{result3}),
which therefore holds for an arbitrary circuit.

\section{Third cumulant contribution}
\label{cumulant}

We have shown in the Sec.\ \ref{PE} that in the long time limit the quantum 
noise contribution to the correlator $J_2(t)$ is small. The same remains true 
for the third cumulant, $J_3(t)$. Since the third cumulant contribution
is small by the parameter ${\cal R}$, we right from the beginning focus on 
its classical part, and rewrite Eq.\ (\ref{J0-3}) as follows: 
\begin{equation}
J_3(t)=(1/6)\langle[\phi(t)-\phi(0)]^3\rangle.
\label{J5}
\end{equation}
Thus we see that $J_3(-t)=-J_3(t)$. This breaks the symmetry between
right and left lead, $P_{LR}\neq P_{RL}$, and the third cumulant adds to the potential
difference across the tunnel junction. In the classical limit $E\ll\Omega$, where
$P(E)$ has a peak, this additional potential simply shifts the energy by a small 
amount $E_3$ that depends on the third cumulant of current: 
$P_{LR}(E)=P(E-E_3)$ and $P_{LR}(E)=P(E+E_3)$. Therefore, the 
function of the symmetric detector (\ref{result3}) has 
to be replaced with 
\begin{equation}
D(\varepsilon,0)=\frac{P_{LR}(\varepsilon)-P_{RL}(-\varepsilon)}{P_{LR}(\varepsilon)+
P_{RL}(-\varepsilon)}=\frac{\varepsilon-E_3}{2\Omega}.
\label{result5}
\end{equation}

The same shift obviously takes place in the asymmetric case. However, there the shift
$E_3$ adds to the shift $E_2$ due to the noise rectification effect (see the discussion
in the end of Sec.\ \ref{classical}). Fortunately, in contrast to the rectification shift,
the energy $E_3$ depends on the direction of current in a mesoscopic system.
Therefore, experimentally the third cumulant contribution can be extracted by changing
the direction of the current through the mesoscopic system. This experimental
technique has been recently used to measure the third cumulant with the help of 
Josephson junction threshold detectors.\cite{threshold-pekola,threshold-huarg} 
In addition, an important role 
in this context plays the universality of the Gaussian noise effect on the detector 
function $D(\varepsilon, 0)$ proven in Sec.\ \ref{universalities}. In what follows we 
evaluate the shift $E_3$ for the cases of fast and slow circuit, 
depending on the circuit response time $\tau_C$.

\subsection{Fast circuit}
\label{fast-cum}

In the case of fast circuit, ${\cal R}\Omega\tau_C\ll 1$, the 
potential fluctuations are Markovian, so that Eq.\ (\ref{J5}) gives
\begin{equation}
J_3(t)=(e^3/6)\l V^3\r t,
\label{J6}
\end{equation}
where $\l V^3\r$ is the Markovian cumulant of the potential. 
According to Refs.\ [\onlinecite{cascade-beenakker,cascade-nagaev,Geneva_group}] 
it is given by $\l V^3\r=R^3\l I^3\r$, where the total third cumulant of the 
current is equal to
\begin{equation}
\l I^3\r=\l I_M^3\r+6\Omega\partial_{V}S(0)+12(\Omega/R)^2\partial_{V}R.
\label{cumulant1}
\end{equation}
Here $\l I_M^3\r$ is the intrinsic third cumulant of the system current, and 
the second and third terms are the ``environmental'' and nonlinear 
cascade corrections, respectively. They originate from the circuit backaction. 

It is useful to write $J_3$ in the form that explicitly shows the coupling 
constant: $J_3(t)=(1/6)(2\pi{\cal R}/e)^3\l I^3\r\, t$. 
We see that indeed such a contribution to the correlator simply shifts the 
energy in the Fourier transform (\ref{PE-expansion}) for the probability 
distribution functions by the amount 
\begin{equation}
E_3=(1/6)(2\pi{\cal R}/e)^3\l I^3\r,\quad {\cal R}\Omega\tau_C\ll 1.
\label{result6}
\end{equation}
In order to estimate the relative effect of the third cumulant we note
that the width of the peak in $P(E)$, where the detector signal is maximum, 
is of the order of ${\cal R}\Omega$, so that $D\sim {\cal R}$.
The energy shift can be estimated as $({\cal R}/e)^3\l I^3\r\sim {\cal R}^2G_MR\Omega$.
Therefore, the relative contribution of the third cumulant is of the order of
${\cal R}G_M/(G_M+G_L)\ll 1$. 

\subsection{Slow circuit}
\label{slow-cum}

In the case of slow circuit, ${\cal R}\Omega\tau_C\gg 1$, the detector ``feels''
slow fluctuations of the potential. Therefore, one can approximate 
$\phi(t)-\phi(0)=e\delta Vt$, according to exact calculations in the Sec.\ \ref{slow}. 
Then the equation (\ref{J5}) gives:
\begin{equation}
J_3(t)=(e^3/6)\langle (\delta V)^3\rangle t^3,
\label{J7}
\end{equation}
where $\langle (\delta V)^3\rangle$ is the third cumulant of equal time 
fluctuations of the potential $V$. In Refs.\ [\onlinecite{Reznikov,ESandAJ}] 
it has been shown that $\langle (\delta V)^3\rangle=(R/3C^2)\l I^3\r$, where the total
current cumulant in this case is given by
\begin{equation}
\l I^3\r=\l I_M^3\r+3\Omega\partial_{V}S(0)+3(\Omega/R)^2\partial_{V}R.
\label{cumulant2}
\end{equation}
It contains the intrinsic cumulant of the system current, $\l I_M^3\r$, and 
the cascade corrections.
Note that in this case, ${\cal R}\Omega\tau_C\gg 1$, the cascade corrections 
are smaller compared the those for a fast circuit, see Eq.\ (\ref{cumulant1}). 
This fact has been recently experimentally
verified in Ref.\ [\onlinecite{Reznikov}]. 

We now substitute the small term (\ref{J7}) to the definition (\ref{PE-expansion}).
This gives $P_{LR}=[1+(e^3/6)\langle (\delta V)^3\rangle\partial^3_E]P(E)$ and
$P_{RL}=[1-(e^3/6)\langle (\delta V)^3\rangle\partial^3_E]P(E)$. Using Eqs.\ (\ref{PE5})
and (\ref{result5}) we obtain
\begin{equation}
E_3=\frac{(2\pi {\cal R})^2}{6\tau_C\Omega e^3}\,\l I^3\r,\quad {\cal R}\Omega\tau_C\gg 1.
\label{result7}
\end{equation}
Note that this energy shift is smaller than the one for the case of fast circuit, 
Eq.\ (\ref{result6}), by the parameter $1/({\cal R}\Omega\tau_C)\ll 1$. 
Since the width of the distribution
$P(E)$ is of the order of $\sqrt{{\cal R}\Omega/\tau_C}$, the relative 
contribution of the third cumulant to the function $D$ can be estimated as 
${\cal R}G_M/[(G_M+G_L)({\cal R}\Omega\tau_C)^{1/2}]\ll 1$.

\section{Outlook}
\label{conclusion}

We have presented the theory of quantum noise detectors based on the resonant tunneling
phenomenon. It is summarized in the introduction, which can also be used as a guide 
to most important results. Here we briefly discuss related problems which yet to be solved.
First of all, it would be interesting to relax the condition of a weak coupling. In the case 
${\cal R}\sim 1$ fluctuation dissipation relations may contain an information about the full
distribution of the fluctuating potential. Interestingly, it has been shown in Ref.\ 
[\onlinecite{sukhorukov2}] that the double dot system in the adiabatic regime, 
$\tau_C\langle I_D\rangle\gg 1$, may serve as a nonlinear element which generates an instability
in the mesoscopic circuit. It then may be used as on-chip threshold detector of rare event 
in transport. The difficulty of this problem is that the dynamical Coulomb blockade effect 
in this case is not generally negligible. 

In the case of a TP detector the excitation of an electron-hole pair in the QPC
may cause a transition in the DD system. This competing quasiparticle process reduces the 
precision of the detection of  collective charge excitations in the measurement circuit. 
Intuitively, one should keep the current through the QPC on a very low level. However, this 
will reduce the rate of the measurement. Moreover, the quasiparticle process is interesting
in itself and should be investigated theoretically.

We think that the physics of double-dot systems described here is rather universal and should 
be same in various two-level systems of a different nature. Nevertheless, it is important to 
consider other systems too. Moreover, it would be interesting to generalize present results 
to the case of a quantum detector with many 
levels with the energies $\varepsilon_n$, $n=1,2,\ldots$. There is a hope that such 
system will be able to detect high order correlators of current at finite frequencies equal
to the energies $\varepsilon_n$.

Concerning specific results presented in the paper, two problems remain to be solved.
First, we have shown that in the classical noise detection regime the effect of the second cumulant 
of the system current is universal, i.e.\ it does not depend on the circuit details.  
On the contrary, the third cumulant contribution depends on the circuit response time and 
has been found here in the limit of fast and slow circuit. Interesting problem, which
may also be experimentally very relevant, is to find the third cumulant contribution including
cascade correction for arbitrary circuit.

Second, the most dramatic effect of a non-equilibrium system noise on the detector function is 
that the exponential behavior (\ref{FDT}) is replaced with the power law functions (\ref{result1})
and (\ref{result1-dimless}). Thus the power law behavior is a signature of non-equilibrium processes.
However, this our result has been obtained by considering a coherent non-interacting mesoscopic 
conductor as an example of the system. Therefore, it would be interesting to consider other systems
in order to check the generality of our conclusion.

\begin{acknowledgments}
We acknowledge the illuminating discussion with the group of Klaus Ensslin.
This work was supported by the Swiss National Science Foundation.

\end{acknowledgments}


\begin{thebibliography}{02}

\bibitem{kane-fisher}
C.L. Kane, and M.P.A. Fisher,
Phys. Rev. Lett. {\bf 72}, 724 (1994).

\bibitem{Glattli}
L. Saminadayar, D.C. Glattli, Y. Jin, and B. Etienne,
Phys. Rev. Lett. {\bf 79}, 2526 (1997).

\bibitem{dePicciotto}
M. Reznikov, R. de Picciotto, T.G. Griffiths, M. Heiblum, and V. Umansky,
Nature (London) {\bf 399}, 238 (1999).


\bibitem{rogovin-scalapino}
D. Rogovin, and D.J. Scalapino, 
Ann. Phys. (N.Y.) {\bf 86}, 1 (1974).

\bibitem{footnote1}
See also the generalization of the theorem to the photon-assisted transport in 
Ref.\ [\onlinecite{sis-tucker}], and to the cotunneling transport in 
Ref.\ [\onlinecite{SBL}].

\bibitem{sis-tucker}
J.R. Tucker, and M.J. Feldman, Rev. Mod. Phys. {\bf 57}, 1055 (1985).

\bibitem{SBL}
E.V. Sukhorukov, G. Burkard, and D. Loss,
Phys. Rev. B {\bf 63}, 125315 (2001);
arXiv:cond-mat/0010458. 


\bibitem{callen-welton}
H.B. Callen, and T.A. Welton, Phys. Rev. {\bf 83}, 34 (1951).

\bibitem{LL03}
D. Landau and E.M. Lifshitz, 
{\em Quantum Mechanics} (Pergamon, London, 1958).


\bibitem{cascade-beenakker}
C.W.J. Beenakker, M. Kindermann, and Yu.V. Nazarov,
Phys. Rev. Let. {\bf 90}, 176802 (2003).

\bibitem{cascade-reulet}
B. Reulet, J. Senzier and D.E. Prober,
Phys. Rev. Let. {\bf 91},
196601 (2003).

\bibitem{Reznikov}
 G. Gershon, Yu. Bomze, E. V. Sukhorukov, M. Reznikov,
 arXiv:cond-mat/0710.1852.

\bibitem{ESandAJ}
E.V. Sukhorukov and A.N. Jordan, Phys. Rev. Lett. {\bf 98},
136803 (2007).

\bibitem{cascade-nagaev}
K.E. Nagaev,
Phys. Rev. B{\bf 66}, 075334 (2002).

\bibitem{Geneva_group}
A.N. Jordan, E.V. Sukhorukov, S. Pilgram,
J. Math. Phys. {\bf 45}, 4386 (2004).

\bibitem{twolevel-fujisawa}
T. Fujisawa, T.H. Oosterkamp, W.G. van der Wiel, 
B.W. Broer, R. Aguado, S. Tarucha, and L.P. Kouwenhoven,
Science {\bf 282}, 932 (1998).

\bibitem{twolevel-aguado}
R. Aguado and L.P. Kouwenhoven, 
Phys. Rev. Lett. {\bf 84}, 1986 (2000).

\bibitem{twolevel-schoelkopf}
R.J. Schoelkopf, A.A. Clerk, S.M. Girvin, K.W. Lehnert, and M.H. Devoret,
in {\em Quantum Noise in Mesoscopic Physics}, edited by Y.V. Nazarov 
(Kluwer, Dordrecht, 2003).

\bibitem{twolevel-onac}
E. Onac, F. Balestro, L.H. Willems van Beveren, U. Hartmann, Y.V. Nazarov, and L.P. Kouwenhoven,
Phys. Rev. Lett. {\bf 96}, 176601 (2006) 

\bibitem{twolevel-khrapai}
V.S. Khrapai, S. Ludwig, J.P. Kotthaus, H.P. Tranitz, and W. Wegscheider,
Phys. Rev. Lett. {\bf 97}, 176803 (2006). 

\bibitem{twolevel-gustavsson}
S. Gustavsson, M. Studer, R. Leturcq, T. Ihn, K. Ensslin, D.C. Driscoll, A.C. Gossard, 
Phys. Rev. Lett. {\bf 99}, 206804 (2007) .

\bibitem{qpc-field}
M. Field {\em et al.}, Phys. Rev. Lett. {\bf 70}, 1311 (1993).

\bibitem{qpc-elzerman}
J.M. Elzerman, R. Hanson1, J.S. Greidanus, L.H. Willems van Beveren, 
S. De Franceschi, L.M.K. Vandersypen, S. Tarucha, and L.P. Kouwenhoven,
Phys. Rev. B {\bf 67}, 161308(R) (2003)

\bibitem{qpc-fujisawa}
T. Fujisawa, T. Hayashi, Y. Hirayama,
H.D. Cheong, and Y. H. Jeong, 
Appl. Phys. Lett. {\bf 84}, 2343 (2004).

\bibitem{qpc-schleser}
R. Schleser, E. Ruh, T. Ihn, K. Ensslin
D.C. Driscoll, and A.C. Gossard,
Appl. Phys. Lett. {\bf 85},
2005 (2004).

\bibitem{qpc-vandersypen}
L.M.K. Vandersypen, J.M. Elzerman, R.N. Schouten, L.H. Willems van Beveren, 
R. Hanson, and L.P. Kouwenhoven, Appl. Phys. Lett. {\bf 85}, 4394 (2004).

\bibitem{charge-dicarlo}
L. DiCarlo, H.J. Lynch, A.C. Johnson, L.I. Childress, K. Crockett, C.M. Marcus,
M.P. Hanson, and A.C. Gossard,
Phys. Rev. Lett. {\bf 92}, 226801 (2004). 

\bibitem{charge-fujisawa}
T. Fujisawa, T. Hayashi, R. Tomita, and Y. Hirayama,
Science {\bf 312}, 1634 (2006).

\bibitem{counting-gustavsson1}
S. Gustavsson, R. Leturcq, B. Simovi, R. Schleser, T. Ihn, P. Studerus, K. Ensslin,
D.C. Driscoll, and A.C. Gossard,
Phys. Rev. Lett. {\bf 96}, 076605 (2006). 

\bibitem{counting-gustavsson2}
S. Gustavsson, R. Leturcq, B. Simovi, R. Schleser, P. Studerus, T. Ihn, K. Ensslin,
D.C. Driscoll, and A. C. Gossard,
Phys. Rev. B {\bf 74}, 195305 (2006).

\bibitem{counting-gustavsson3}
S. Gustavsson, R. Leturcq, T. Ihn, K. Ensslin, M. Reinwald, W. Wegscheider, 
Phys. Rev. B {\bf 75}, 075314 (2007) 

\bibitem{sukhorukov1}
E.V. Sukhorukov, A.N. Jordan, S. Gustavsson, R. Leturcq, Th. Ihn, and K. Ensslin, 
Nature Physics {\bf 3}, 243 (2007). 

\bibitem{threshold-pekola}
A.V. Timofeev, M. Meschke, J.T. Peltonen, T.T. Heik\-kil\"a, and J.P. Pekola, 
Phys. Rev. Lett. {\bf 98}, 207001 (2007).

\bibitem{threshold-huarg}
B. Huard, H. Pothier, N.O. Birge, D. Est\`eve, X. Waintal, and
J. Ankerhold, Ann. Phys. (Leipzig) {\bf 16}, 736 (2007).

\bibitem{ingold-nazarov}
G.-L. Ingold and Y.V. Nazarov, in {\em Single
Charge Tunneling}, edited by H. Grabert and M.H. Devoret
(Plenum, New York, 1992), Chap. 2.

\bibitem{footnote2}
The stationarity $\langle\phi^2(t)\rangle=\langle\phi^2(0)\rangle$, often used in literature,
\cite{ingold-nazarov} breaks down in a long time limit. Thus, we do not rely on this approximation.

\bibitem{asymmetry-lesovik}
G.B. Lesovik, and R. Loosen, 
JETP lett. 65, 295 (1997).

\bibitem{asymmetry-gavish1}
U. Gavish, Y. Levinson, and Y. Imry
Phys. Rev. B {\bf 62}, R10637 (2000).

\bibitem{asymmetry-gavish2}
U. Gavish, Y. Imry, Y. Levinson, and B. Yurke 
in {\em Quantum Noise in Mesoscopic Physics},
edited by Y.V. Nazarov (Kluwer, Dordrecht, 2003).

\bibitem{blockade-altimiras}
For a recent experiment,see C. Altimiras, U. Gennser, A. Cavanna, D. Mailly, F. Pierre, 
Phys. Rev. Lett. {\bf 99}, 256805 (2007). 

\bibitem{jonathan}
J. Edwards, Master thesis (Geneva, June 2007).

\bibitem{harris}
R. Harris {\em et al.}, 
arXiv:0712.0838v2.  

\bibitem{averin}
M.H.S. Amin, D.V. Averin,
arXiv:0712.0845. 

\bibitem{footnote3}
In general, the phase $\phi$ is multiplied by a dimensionless number parameterizing 
the efficiency of screening by a circuit capacitor. This number
merely renormalizes the coupling constant which does not affect the  
the detector function. 

\bibitem{blanter-buttiker}
For a review, see Ya. M. Blanter and M. B\"uttiker, 
Physics Reports {\bf 336}, 1-166 (2000).

\bibitem{sukhorukov2}
A.N. Jordan, E.V. Sukhorukov,
Phys. Rev. B {\bf 72}, 035335 (2005). 


\end{thebibliography}
\end{document}